\documentclass[aps,prl,reprint,superscripraddress]{revtex4-1}
\usepackage{amsmath,amsfonts,amssymb,bm}
\usepackage{graphicx}
\usepackage{color}
\usepackage{subfigure}
\usepackage{multirow}
\usepackage{textcomp}
\usepackage{slashed}

\definecolor{purple}{rgb}{0.5,0,0.5}
\definecolor{blue}{rgb}{0.0,0,0.9}

\usepackage[colorlinks=true, pdfstartview=FitV, linkcolor=purple, citecolor= purple, urlcolor=blue]{hyperref}

\begin{document}


\title{Excited $B_{c}$ States via Continuum QCD}




\author{Lei Chang}\email{leichang@nankai.edu.cn}
\affiliation{School of Physics, Nankai University, Tianjin 300071, China}

\author{Muyang Chen}\email{muyang@nankai.edu.cn}
\affiliation{School of Physics, Nankai University, Tianjin 300071, China}

\author{Yu-xin Liu}\email{yxliu@pku.edu.cn}
\affiliation{Department of Physics and State Key Laboratory of Nuclear Physics and Technology, Peking University, Beijing 100871, China}
\affiliation{Collaborative Innovation Center of Quantum Matter, Beijing 100871, China}
\affiliation{Center for High Energy Physics, Peking University, Beijing 100871, China}

\date{\today}

\begin{abstract}
%
We study the most recently observed excited $B_{c}$ states with the Dyson--Schwinger equation and the Bethe--Salpeter equation approach of continuum QCD. 
The obtained $M_{B^+_{c}(2S)}=6.813(16)\textmd{ GeV}$, $M_{B^{*+}_{c}(2S)}=6.841(18)\textmd{ GeV}$ and the mass splitting $M_{B_c^+(2S)}-M^{\text{rec}}_{B_c^{*+}(2S)}  \approx 0.039 \textmd{ GeV}$ 
%
%
agree with the observations very well. Moreover we predict the leptonic decay constant $f_{B^+_{c}(2S)}=-0.165(10)\textmd{ GeV}$, $f_{B^{*+}_{c}(2S)}=-0.161(7)\textmd{ GeV}$ respectively.

\end{abstract}
%

\maketitle


 \noindent\emph{1.\,Introduction}\,---\,
Recently, two excited $B_c$ mesons, $B^+_{c}(2S)$ and $B^{*+}_{c}(2S)$, were observed in the mass spectrum of $B_c^+\pi^+\pi^-$ for the first time
by the CMS experiment at $\sqrt{s}=13\textmd{ TeV}$~\cite{Sirunyan:2019osb}.
The mass of $B^+_{c}(2S)$ is determined to be $M_{B^+_{c}(2S)} = 6871.0 \pm 1.2(stat.) \pm 0.8(syst.) \pm 0.8(B_c^+)$ MeV,
while the mass difference $M_{B^+_{c}(2S)} - M^{\text{rec}}_{B^{*+}_{c}(2S)} = 29.0 \pm 1.5(stat.) \pm 0.7(syst.)$ MeV, where $M^{\text{rec}}_{B^{*+}_{c}(2S)}$ is defined as $M^{\text{rec}}_{B^{*+}_{c}(2S)} = M_{B^{*+}_{c}(2S)} - (M_{B^{*+}_{c}(1S)}-M_{B^+_{c}(1S)})$.
The above results are then confirmed by the LHCb experiment with 8.5 fb$^{-1}$ $pp$ collision data~\cite{LHCb},
being $M_{B^+_{c}(2S)} = 6872.1 \pm 1.3(stat.) \pm 0.1(syst.) \pm 0.8(B_c^+)$ MeV and $M_{B^+_{c}(2S)} - M^{\text{rec}}_{B^{*+}_{c}(2S)} = 31.0 \pm 1.4(stat.)$ MeV, respectively.

Investigating the open flavor states such as the $B_{c}^+$ family of $(c\bar{b})$ mesons could enrich our understanding of the strong interaction.
There have been plenty of theory studies and we refer to Ref.~\cite{Eichten:2019gig} and the references therein for the contemporary statements.
Exploring the excited states relies on the detailed understanding of long range behavior of strong interaction and encounters the difficulties due to the intrinsic complexity.
The quark model has been thoroughly applied to study hadron spectrum and, by using a phenomenal nonrelativistic potential model, the mass spectrum and decay properties of $(c\bar{b})$ mesons have also been explored(see, e.g. Ref.~\cite{Soni:2017wvy}). However, investigations based on {\it ab initio} theory of strong interactions, Quantum Chromodynamics(QCD)(QCD), are still challenges. The precise predictions of charmed-bottom ground state from Lattice QCD(lQCD)~\cite{Mathur:2018epb} has been released recently with the the masses $M_{B_c^+} = 6276(3)(6)$MeV and $M_{B_c^{+*}}= 6331(4)(6)$MeV respectively.  
Studying the masses of the excited states in lQCD are more difficult than determing those of the ground states accurately~\cite{Dudek:2007wv,Burch:2009az} and the leptonic decay constants of excited $B_{c}^+$ states have not yet been touched.
For details of the difficulties to study the decay constant in lQCD simulations please refer to Refs.~\cite{McNeile:2006qy, Mastropas:2014fsa}, where trials of calculating the decay constant of the first radial excited pion are carried out with the inspiration of a continuum theory prediction~\cite{Holl:2004fr}.

As a continuum functional method of QCD, the Dyson-Schwinger equation and Bethe-Salpeter equation (DSBSE)~\cite{Roberts:1994dr, Maris:2003vk, Bashir:2012fs} approach is complementary to lQCD and a covariant way to bridge the hadron physics and the fundamental degree of QCD. The difficulty of investigating the open flavor hadrons within this approach has been reported in Ref.~\cite{Nguyen:2010yh} and that for exotic and radial excited states has been displayed in Ref.~\cite{Qin:2011xq}. Then some efforts (for example Ref.~\cite{Ivanov:1998ms, Blank:2011ha, Fischer:2014cfa}) have been made.  Using an algebraic model, the mass of $B_{c}^{*+}$,  which is consistent with the world average value, has been predicted~\cite{Gomez-Rocha:2016cji}. However it is not possible to predict the decay constants and the properties of the radial excited states in that framework, because the interaction lacks the relative momentum dependence. A novel extrapolation method has been developed in Ref.~\cite{Binosi:2018rht}. Therein the obtained masses and decay constants of the ground states mesons are comparable to experimental measurements and lQCD simulations, showing the success of the rainbow ladder (RL) approximation. What's more, taking into account the flavor dependence of the quark-gluon interaction properly, we give a successful and unified description of the ground states of the open flavor mesons and the quarkonia~\cite{Chen:2019otg}. Our results of the heavy mesons deviate from the experiment and lQCD results only about $1\%$ for the ground state masses and less than $7\%$ for the decay constants. The predicted masses of the $B_c$ mesons, $M_{B_c^+} = 6290(3)$ MeV and $M_{B_c^{*+}}= 6357(4)$ MeV, are comparable with the experimental and lQCD values.

To study the excited states of $B_{c}^{+}$ and $B_{c}^{*+}$ in the continuum QCD approach directly, one should develop a scheme by extending those given in Refs.~\cite{Binosi:2018rht} and \cite{Chen:2019otg}. In the extension, one should maintain the parameters as the same as (without any fine tuning) the ones which produce the masses and decay constants of the ground states successfully. In this Letter, we produce the masses and the decay constants of the first excited states, $B_{c}^{+}(2S)$ and $B_{c}^{*+}(2S)$, via the continuum QCD approach. Our obtained mass of the excited states agree with the experimental observations very well. The obtained decay constants are also quite reasonable.

\medskip

\noindent\emph{2.\,DSBSE approach}\,---\,
Here we present the RL truncated DSBSE approach which takes into account the flavor dependence of the quark-gluon interaction properly~\cite{Chen:2019otg}.
The BS equation is
\begin{equation}\label{eq:BSE}
  \Gamma^{fg}(k;P)  =
 -\frac{4}{3}[Z_{2}]^{2} \int^\Lambda_{d q} \big{[} D^{fg}_{\mu\nu}(k-q)\gamma_{\mu}^{} \chi^{fg}(q;P) \gamma_{\nu}^{} \big{]} ,
\end{equation}
where $f$ and $g$ label the quark flavor, $\Gamma^{fg}(k;P)$ is the Bethe-Salpeter amplitude (BSA), $k$ and $P$ are the relative and total momentum of the meson.
$\chi^{fg}(q;P) = S_{f}(q_{+}) \Gamma^{fg}(q;P) S_{g}(q_{-})$ is the BS wave function, $S_{f}(q_{+})$ and $S_{g}(q_{-})$ are the quark propagators,
where $q_{+} = q + \iota P/2$, $q_{-} = q - (1-\iota) P/2$, $\iota$ is the partitioning parameter describing the momentum partition between quark and antiquark
and doesn't affect the physical observables.
The quark propagators satisfy the DS equation,
\begin{eqnarray}\nonumber
S_f^{-1}(p) &=& Z_2 (i\gamma\cdot p + Z_m m_f) \\\label{eq:DSE}
&&+ \frac{4}{3}[Z_2]^2 \int^\Lambda_{d q} D^{ff}_{\mu\nu}(p-q)\gamma_\mu S_{f}(q)\gamma_\nu.
\end{eqnarray}
In Eq.(\ref{eq:BSE}) and Eq.(\ref{eq:DSE}), $\int^\Lambda_{d q}=\int ^{\Lambda} d^{4} q/(2\pi)^{4}$ stands for a Poincar$\acute{\text{e}}$ invariant regularized integration, with $\Lambda$ the regularization mass-scale,
$m_f$ is the current quark mass at renormalization scale $\zeta$, $Z_2$ and $Z_m$ are the renormalization constants of the quark field and the quark mass depending on $\Lambda$ and $\zeta$.
We adopt a flavor independent renormalization scheme and choose $\zeta=2\textmd{ GeV}$.
$D^{fg}_{\mu\nu}(l) = \left(\delta_{\mu\nu}-\frac{l_{\mu}l_{\nu}}{l^{2}}\right)\mathcal{G}^{fg}(l^2)$ is the gluon propagator including the effect of the flavor dependence of the dressed quark-gluon-vertex.
The dressed function $\mathcal{G}^{fg}(s)$ is composed of a flavor dependent infrared(IR) part and a flavor independent ultraviolet(UV) part,
\begin{eqnarray}\label{eq:gluonfmodel}
  \mathcal{G}^{fg}(s) 		&=& \mathcal{G}_{IR}^{fg}(s) + \mathcal{G}_{UV}(s),\\\label{eq:gluonfInfrared}
  \mathcal{G}_{IR}^{fg}(s) 	&=& 8\pi^2\frac{D_f}{\omega_f^2}\frac{D_g}{\omega_g^2} e^{-s/(\omega_f\omega_g)},\\\label{eq:gluonfUltraviolet}
  \mathcal{G}_{UV}(s) 		&=& \frac{8\pi^{2} \gamma_{m}^{} \mathcal{F}(s)}{\text{ln}[\tau+(1+s/\Lambda^{2}_{QCD})^2]},
\end{eqnarray}
where $\mathcal{F}(s)=[1 - \exp(-s^2/[4m_{t}^{4}])]/s$, $\gamma_{m}^{}=12/(33-2N_{f})$, with $m_{t}=1.0 \textmd{ GeV}\,$, $\tau=e^{10} - 1$, $N_f=5$, and $\Lambda_{\text{QCD}}=0.21 \textmd{ GeV}\,$.

In Eq.(\ref{eq:gluonfInfrared}), $D_{f,g}$ and $\omega_{f,g}$ are parameters expressing the flavor dependent quark-gluon interaction, which are fixed by physical observables.
Three groups of parameters corresponding to a varying of the interaction width are given in Ref.~\cite{Chen:2019otg}.
The parameters of the charm and beauty system are listed in Table~\ref{tab:parametersBC}.
The current mass on the mass shell is defined by
\begin{eqnarray}
 \check{m}_f^{\zeta} &=& \left.\hat{m}_f\middle/\left(\frac{1}{2}\textmd{ln}\frac{\check{m}^2}{\Lambda^2_{\textmd{QCD}}}\right)^{\gamma_m}\right.,\\
 \hat{m}_f	 &=& \lim_{p^2 \to \infty}\left(\frac{1}{2}\textmd{ln}\frac{p^2}{\Lambda^2_{\textmd{QCD}}}\right)^{\gamma_m} M_f(p^2),
\end{eqnarray}
where $\hat{m}_f$ is the renormalisation-group invariant current-quark mass\cite{Maris:1997tm} and $M_f(p^2)$ is the quark mass function in the quark propagator $S_f(p)=\frac{Z_f(p^2,\zeta^2)}{i\gamma\cdot p + M_f(p^2)}$.
We extract the value $\check{m}_{c}=1.31 \textmd{ GeV}$ and $\check{m}_{b}=4.27 \textmd{ GeV}$, which are commensurate with those given by PDG~\cite{Tanabashi:2018oca}.

\begin{table}[t!]
\caption{\label{tab:parametersBC} Three groups of parameters quoted from Ref.\cite{Chen:2019otg}. $\omega_f$ and $D_f$ are both measured in GeV.}
\begin{tabular}{c|c|c|c|c|c|c|c|c|c}
\hline \hline
flavor& \hspace*{-0.1cm}&\; $\omega_f $\; &\; $D_f^2$ \;&\hspace*{-0.1cm}&\; $\omega_f $\; &\; $D_f^2$ \;	 &\hspace*{-0.1cm}	&\; $w_f $\; &\; $D_f^2$ \;\\ [0.5mm]
\hline
$c$	&\hspace*{-0.1cm}	& 0.690 & 0.645 		&\hspace*{-0.1cm}	& 0.730 & 0.599 		&\hspace*{-0.1cm}	& 0.760 & 0.570 \\
$b$	&\hspace*{-0.1cm}	& 0.722 & 0.258 		&\hspace*{-0.1cm}	& 0.766 & 0.241 		&\hspace*{-0.1cm}	& 0.792 & 0.231 \\
\hline \hline
\end{tabular}
\end{table}

\medskip

\noindent\emph{3.\,Extrapolation}\,---\,
The quark propagators in Eq.(\ref{eq:BSE}) are functions of the complex momenta $q^2_{\pm}$ which lies in a parabolic region.
Any singular structure in the quark propagator indicates the upper bound of the maximum bound state mass obtainable directly, $P^2>-M^2_{max}$, where $-M^2_{max}$ defines the contour border of the parabolic region.
Due to color confinement, the quark propagators indeed have such singularities.
The existing nodes in the Schwinger function of the charm and bottom quark propagators reveal the information of the complex conjugate poles~\cite{Chen:2016bpj, Roberts:2007ji}.
The ground states are within the parabolic region and the masses and the leptonic decay constants can be obtained directly.
However, the radial excited states are outside the parabolic region, hence we adopt an extrapolation scheme to determine the masses and the decay constants.

The BSE can be viewed as a $P^2$-dependent eigenvalue problem,
\begin{equation}\label{eq:BSE2}
  \lambda^{fg}(P^2) \big{[} \Gamma^{fg}(k;P)  \big{]}^{\alpha}_{\beta}  =   \int^\Lambda_{d q} \big{[} K(k,q;P) \big{]}^{\alpha\delta}_{\sigma\beta} \big{[} \chi^{fg}(q;P)  \big{]}^{\sigma}_{\delta} ,
\end{equation}
where $\big{[} K(k,q;P) \big{]}^{\alpha\delta}_{\sigma\beta} = -\frac{4}{3}[Z_{2}]^{2} D^{fg}_{\mu\nu}(k-q) [\gamma_{\mu}^{}]^{\alpha}_\sigma [\gamma_{\nu}]^\delta_\beta$,
and $\alpha$, $\beta$, $\sigma$ and $\delta$ are the Dirac indexes.
The meson mass is determined by $\lambda^{fg}(P^2=-M^2)=1$. An extrapolation to the physical bound state mass should be implemented while the state mass is larger than the contour border $M^2_{max}$. We use a Pad$\acute{\text{e}}$ approximation
\begin{equation}\label{eq:fitlambda}
 \frac{1}{\lambda^{fg}(P^2)} = \frac{1 + \sum^{\infty}_{n=1}\, a_n (P^2+s)^n}{1 + \sum^{\infty}_{n=1}\, b_n (P^2+s)^n},
\end{equation}
to fit the $\lambda^{fg}(P^2)$, with $s$, $a_n$ and $b_n$ the parameters.
The leptonic decay constant of the pseudoscalar meson ($0^-$) and vector meson ($1^-$) are defined by
\begin{eqnarray}\label{eq:definition-f0-}
 f^{fg}_{0^{-}}(P^2)P_{\mu} &=& Z_{2} N_{c} \;\text{tr} \! \int^{\Lambda}_{d k} \! \gamma_{5}^{} \gamma_{\mu}^{} \chi^{fg}_{0^{-}}(k;P), \\\label{eq:definition-f1-}
 f^{fg}_{1^{-}}(P^2) \sqrt{-P^{2}}&=& \frac{Z_{2} N_{c}}{3} \;\text{tr} \! \int^{\Lambda}_{d k} \!  \gamma_{\mu}^{} \chi^{fg}_{1^{-},\mu}(k;P),
\end{eqnarray}
with $\text{tr}$ the trace of the Dirac index.
$f^{fg}(P^2)$ is generally fitted by
\begin{equation}\label{eq:fitf}
 f^{fg}(P^2) = \frac{f_0 + \sum^{\infty}_{n=1}\, c_n (P^2+s)^n}{1 + \sum^{\infty}_{n=1}\, d_n (P^2+s)^n},
\end{equation}
where $f_0$, $c_n$ and $d_n$ are parameters, and $s=M^2$ is the square of the mass.
The physical decay constant is $f^{fg}(-M^2) = f_0$.

\begin{figure}[t!]
 \includegraphics[width=0.45\textwidth]{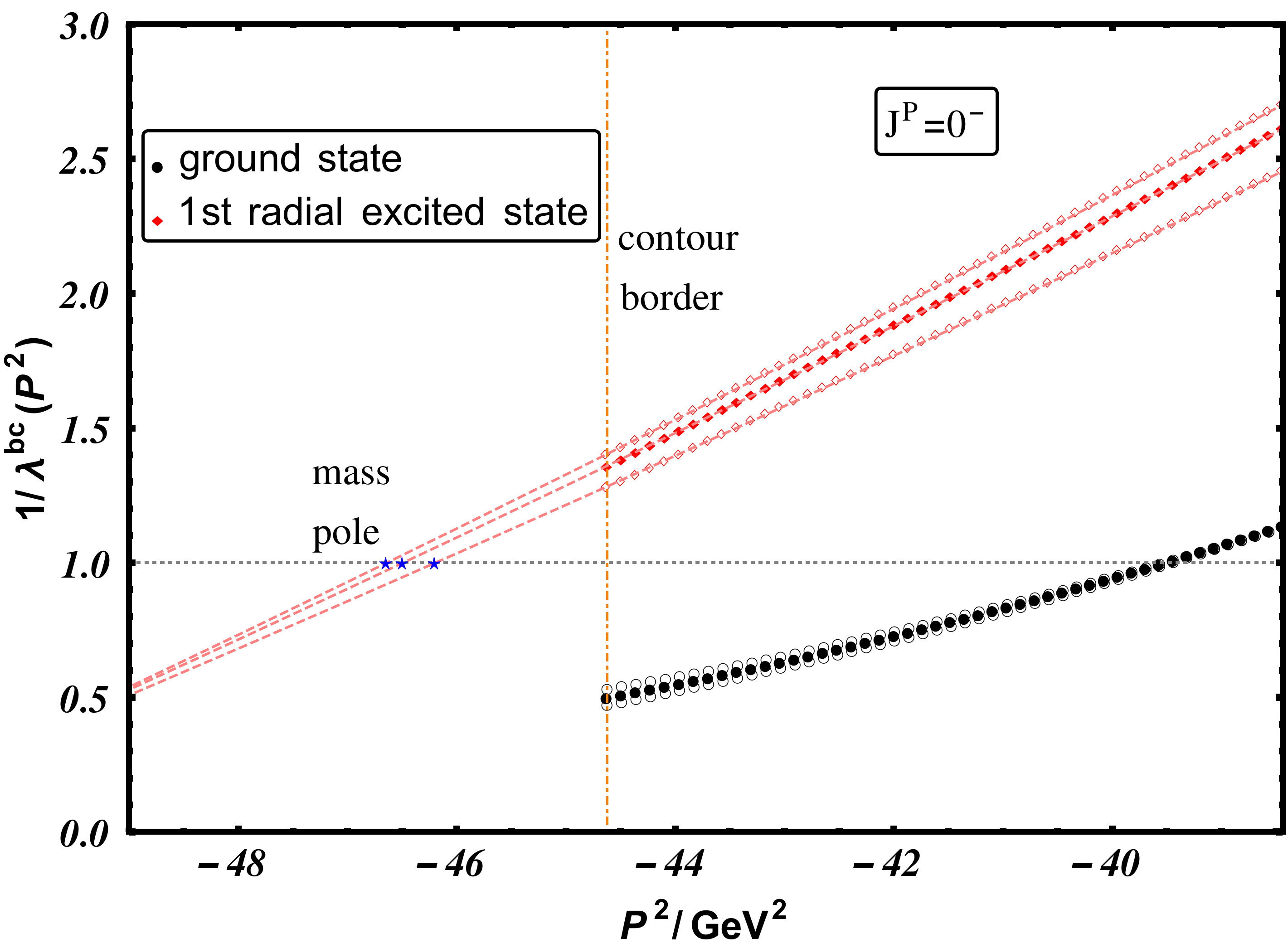}
\caption{\label{fig:lambdaPP}
$P^{2}$ dependence of $1/\lambda^{bc}$ for $J^P=0^-$ state.
The black circles correspond to the ground state, and red diamonds correspond to the first radial excited state.
The open circles and diamonds correspond to the varying of the parameters in Table.\ref{tab:parametersBC}.
The virtical dot-dashed is the contour border on the right of which the direct calculation can be applied.
The blue stars present our extrapolated first radial excited state.}
\end{figure}

\begin{table}[t!]
\caption{\label{tab:mass} Masses of the first radial excited states of charm-beauty system (in GeV).
The experimental data for $M_{\eta_{c}(2S)}$, $M_{\psi(2S)}$, $M_{\eta_{b}(2S)}$ and $M_{\varUpsilon(2S)}$ are taken from Ref.\cite{Tanabashi:2018oca},
$M_{B^+_{c}(2S)}$ and $M_{B^+_{c}(2S)}-M^{\text{rec}}_{B^{*+}_{c}(2S)}$ from Ref.\cite{LHCb}. The mass splitting, $M_{B^{*+}_{c}(1S)}-M_{B^+_{c}(1S)}$, is quoted from Ref.\cite{Chen:2019otg}.
The uncertainties of our results correspond to the varying of the parameters in Table~\ref{tab:parametersBC}.}
\begin{tabular}{c|c|c|c}
\hline
	&$M_{\eta_{c}(2S)}$	&$M_{\psi(2S)}$	&$M_{\psi(2S)}-M_{\eta_{c}(2S)}$	\\
\hline
here	&$3.606(18)$		&$3.645(18)$	&0.039					\\
expt.	&$3.638(1)$		&$3.686(1)$	&0.048					\\
\hline\hline
	&$M_{B^+_{c}(2S)}$	&$M_{B^{*+}_{c}(2S)}$	&$M_{B^+_{c}(2S)}-M^{\text{rec}}_{B^{*+}_{c}(2S)}$	\\
\hline
here	&$6.813(16)$		&$6.841(18)$	&0.039					\\
expt.	&$6.872(2)$		&--		&0.031					\\
\hline\hline
	&$M_{\eta_{b}(2S)}$	&$M_{\varUpsilon(2S)}$	&$M_{\varUpsilon(2S)}-M_{\eta_{b}(2S)}$	\\
\hline
here	&$9.915(15)$		&$9.941(15)$	&0.026					\\
expt.	&$9.999(4)$		&$10.023(1)$	&0.024					\\
\hline
\end{tabular}
\end{table}

\medskip

\noindent\emph{4.\,Results}\,---\,
The series Eq.(\ref{eq:fitlambda}) converges very fast, a good fitting is obtained for $n=1$.
An illustration of the mass extrapolation is given by Fig.~\ref{fig:lambdaPP}, which is the case of $B^+_{c}$.
The black circles show the $1/\lambda^{bc}(P^2)$ of the ground state $B^+_{c}(1S)$.
The mass, $M_{B^+_{c}(1S)}$, lies in the parabolic region defined by the singularities of the quark propagator, so it is obtained directly.
The red diamonds show the $1/\lambda^{bc}(P^2)$ of the first radial excited state $B^+_{c}(2S)$.
$M_{B^+_{c}(2S)}$ lies outside the parabolic region, and its value is extrapolated and presented by the blue stars.
The open circles and diamonds correspond to the varying of the parameters in Table~\ref{tab:parametersBC}, which is the main uncertainty of our results. The other excited states are analysis by the similar method.

The masses of the first radial excited state of the charm-beauty system are listed in Table~\ref{tab:mass}. The average of the results with the three sets of parameters is quoted as final result and the uncertainties are set from the difference between the average and the largest and smallest value respectively.  The excited meson masses increase with the value of parameter $\omega$ showing more sensitive than the ground state as being pointed out by others (see, e.g., Ref.~\cite{Blank:2011qk}).
The relative errors of our results to the experimental date are within $1\%$.
What's more, the mass differences of the vector meson and pseudoscalar meson, $M_{\psi(2S)}-M_{\eta_{c}(2S)}$ and $M_{\varUpsilon(2S)}-M_{\eta_{b}(2S)}$, are comparable with the experimental value.
The reconstructed masses are defined by
\begin{equation}
M^{\text{rec}}_{B^{*+}_{c}(2S)} = M_{B^{*+}_{c}(2S)}-(M_{B^{*+}_{c}(1S)}-M_{B^+_{c}(1S)}).
\end{equation}
The mass splitting, $M_{B^+_{c}(2S)}-M^{\text{rec}}_{B^{*+}_{c}(2S)}$, is consistent with the recent measurement~\cite{LHCb}.
There is no experimental measurements of $M_{B^{+*}_{c}(1S)}$ and $M_{B^{+*}_{c}(2S)}$ hitherto, our predication waits for the future experimental verification.

To first order in the violation of unitary symmetry, the masses obey the equal spacing rule~\cite{Okubo:1961jc,GellMann:1962xb}:
\begin{eqnarray}\label{eq:srule0-}
 (M_{\eta_{c}(2S)} + M_{\eta_{b}(2S)})/2 &=& M_{B^+_{c}(2S)},\\\label{eq:srule1-}
 (M_{\psi(2S)} + M_{\varUpsilon(2S)})/2 &=& M_{B^{*+}_{c}(2S)}.	
\end{eqnarray}
Our results show that the two sides of Eq.(\ref{eq:srule0-}) and Eq.(\ref{eq:srule1-}) differ by only $0.05\textmd{ GeV}$ which is also consistent to the proposal of the mass inequality in Ref.~\cite{Witten:1983ut}.

\begin{figure}[h!]
 \includegraphics[width=0.45\textwidth]{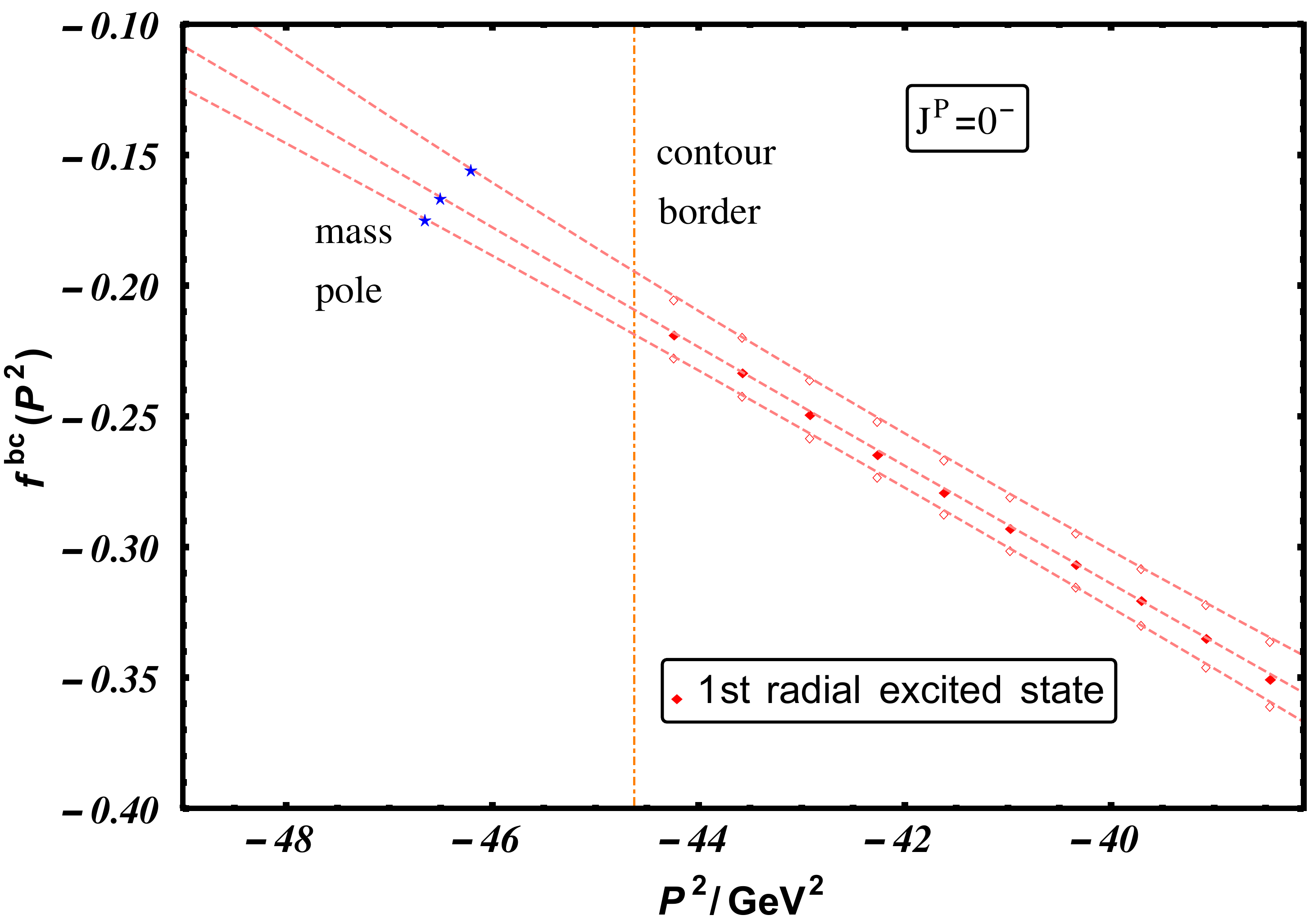}
\caption{\label{fig:fPP}
$P^{2}$ dependence of $f^{bc}$ for the first radial excited $J^P=0^-$ state.
The open diamonds correspond to the varying of the parameters in Table.\ref{tab:parametersBC}.
The virtical dot-dashed is the contour border on the right of which the direct calculation can be applied.
The blue stars present our extrapolated value.}
\end{figure}

The series Eq.~(\ref{eq:fitf}) for the leptonic decay constants also converges very fast, a good fitting is obtained also for $n=1$.
An illustration of the extrapolation of the decay constants is given in Fig.~\ref{fig:fPP}, 
which is the case of $B^+_{c}$.
The physical value is extrapolated and presented by the blue stars. 
Our predication of the decay constants of the first radial excited beauty charmed mesons are listed in Table~\ref{tab:f}. 
We estimate the uncertainty by the similar method as the mass extrapolation.
\begin{table}[h!]
\caption{\label{tab:f} Our predications of the decay constants of the first radial excited beauty charmed mesons (in GeV).
The uncertainties correspond to the varying of the parameters in Table~\ref{tab:parametersBC}.}
\begin{tabular}{c|c|c|c|c|c}
\hline
$f_{\eta_{c}(2S)}$	&$f_{\psi(2S)}$	&$f_{B^+_{c}(2S)}$	&$f_{B^{*+}_{c}(2S)}$	&$f_{\eta_{b}(2S)}$	 &$f_{\varUpsilon(2S)}$	\\
\hline
	-0.097(2)	&-0.119(6)	&-0.165(10)		&-0.161(7)		&-0.310(5)		&-0.320(6)		\\
\hline
\end{tabular}
\end{table}
There is some suppressions for the absolute value of decay constant of excited state comparing to ground state case which agrees with the previous findings~\cite{Holl:2004fr,Holl:2004un, Holl:2005vu, Krassnigg:2016hml} and the difference between the excited and ground states decreases with the increasing of the meson mass.

\medskip

\noindent\emph{5.\,Conclusion}\,---\,
Very recently, CMS and LHCb reported the observation of two excited $B_{c}$ states with high precision~\cite{Sirunyan:2019osb,LHCb}.  Although they are the {\it normal} states within the quark model language, the authors claim that the precision measurements open up an opportunity for the study of hadron physics based on the {\it ab initio} theory of strong interactions. 
In this work, making use of a scattering kernel expressing the flavor dependent quark-gluon interaction properly which describes the ground pseudoscalar and vector mesons successfully,
we produce for the first time the masses and the leptonic decay constants of the first radial excited beauty charmed mesons, $B^+_c(2S)$ and $B^{*+}_c(2S)$, in a continuum QCD directly. The obtained masses are consistent with the recent observations of CMS and LHCb collaborations and the mass splitting $M_{B_c^+(2S)}-M^{\text{rec}}_{B_c^{*+}(2S)}$ is comparable with the experimental result.
The obtained masses of the beauty-charm system also satisfy the equal spacing rule relation approximately.
Furthermore the predicted leptonic decay constants may shed light on the future experimental detection.

\bigskip
\section*{Acknowledgments}
We acknowledge helpful conversations with Pianpian Qin, Sixue Qin, Craig Roberts and Minggang Zhao.
This work is supported by: the Chinese Government Thousand Talents Plan for Young Professionals and the National Natural Science Foundation of China under contracts No. 11435001, and No. 11775041, the National Key Basic Research Program of China under contract No. 2015CB856900.


\end{document}